\documentstyle[12pt]{article}

\begin{document}

\begin{center}
Applications of methods of random differential geometry 
to quantum statistical systems
\end{center}
\begin{center}
Maciej M. Duras
\end{center}
\begin{center}
Institute of Physics, Cracow University of Technology, 
ulica Podchor\c{a}\.zych 1, PL-30084 Cracow, Poland
\end{center}

\begin{center}
Email: mduras @ riad.usk.pk.edu.pl
\end{center}

\begin{center}
AD 2002 May 25
\end{center}

\begin{center}
"International Conference on Theoretical Physics (TH-2002)";
22 July 2002 - 27 July 2002;
Paris, France, UNESCO; 
Theme 9: Session on Random Matrices and Applications.
\end{center}

\section{Abstract}
\label{sec-Abstract}
We apply concepts of random differential geometry connected to the 
random matrix ensembles of the random linear operators acting on 
finite dimensional Hilbert spaces. The values taken by random linear operators 
belong to the Liouville space. This Liouville space is endowed 
with topological and geometrical random structure. 
The considered random eigenproblems for the operators are applied 
to the quantum statistical systems. 
In the case of random quantum Hamiltonians we study both hermitean 
(self-adjoint) and non-hermitean (non-self-adjoint) operators 
leading to Gaussian and Ginibre ensembles Refs. [1], [2], [3].

[1] M. M. Duras, 
``Finite-difference distributions for the Ginibre ensemble,'' 
{\em J. Opt. B: Quantum Semiclass. Opt.} {\bf 2}, 287-291, 2000.

[2] M. M. Duras, K. Sokalski, 
``Finite Element Distributions in Statistical Theory of Energy Levels 
in Quantum Systems,'' 
{\em Physica} {\bf D125}, 260-274, 1999.

[3] M. M. Duras, K. Sokalski, 
``Higher-order finite-element distributions in statistical theory 
of nuclear spectra,''
{\em Phys. Rev. E } {\bf 54}, 3142-3148, 1996. 

\section{Introduction}
\label{sec-Introduction}
We study generic quantum statistical systems with energy dissipation.
Let us consider Hilbert's space $\cal{V}$ with some basis
$\{| \Psi_{i} \rangle\}$.
The space of the linear bounded operators $\hat{X}$ 
acting on Hilbert space $\cal{V}$
is called Liouville space and is denoted $\cal{L}(\cal{V})$.
The Liouville space is again Hilbert's space with the scalar product:
$\langle \hat{X} | \hat{Y} \rangle = {\rm Tr} (\hat{X}^{\dagger} \hat{Y})$.
Hence, it is a Banach space with norm:
$|| \hat{X} || = (\langle \hat{X} | \hat{X} \rangle)^{1/2}$,
and it is metric space with distance:
$\rho (\hat{X}, \hat{Y})= || \hat{X} - \hat{Y} ||$.
Finally, it is topological space with the balls
$B(\hat{X}, r)=\{ \hat{Y}| \rho (\hat{Y}, \hat{X}) < r\}$.
The Liouville space is also a differentiable manifold.
The exist a tangent space ${\rm T}_{\hat{X}}\cal{L}(\cal{V})$,
tangent bundle ${\rm T}\cal{L}(\cal{V})$,
cotangent space ${\rm T}_{\hat{X}}^{\star}\cal{L}(\cal{V})$,
and cotangent bundle ${\rm T}^{\star}\cal{L}(\cal{V})$.  
The quantum operator $\hat{X} \in \cal{L}(\cal{V})$ is in the given
basis a matrix with elements $X_{ij}$.
We are allowed to define random operator variable
$\hat{\cal{X}}: \Omega \ni \omega 
\rightarrow \hat{\cal{X}}(\omega) \in \cal{L}(\cal{V}),$
where $\Omega$ is sample space, and $\omega$ is sample point.
In the basis it is reduced to random matrix variable
${\cal{X}}: \Omega \ni \omega 
\rightarrow {\cal{X}}(\omega) \in {\rm MATRIX}(N,N,{\bf F}),$
where ${\rm MATRIX}(N,N,{\bf F})$ is set of all $N \times N$
matrices with elements from the field ${\bf F}$
\cite{Haake 1990,Guhr 1998,Mehta 1990 0}.
Thus, we define random Hamiltonian operator variable
$\hat{\cal{H}}: \Omega \ni \omega 
\rightarrow \hat{\cal{H}}(\omega) \in \cal{L}(\cal{V}),$
and we define random Hamiltonian matrix variable 
${\cal{H}}: \Omega \ni \omega 
\rightarrow {\cal{H}}(\omega) \in {\rm MATRIX}(N,N,{\bf F})$.
Let us assume that the Hamiltonian $\cal{H}$ is not hermitean operator, thus
its eigenenergies $Z_{i}$ are
complex-valued random variables.
We assume that distribution of matrix elements $H_{ij}$
is governed by Ginibre ensemble
\cite{Haake 1990,Guhr 1998,Ginibre 1965,Mehta 1990 1}.
$\cal{H}(\omega)$ belongs to general linear Lie group GL($N$, {\bf C}),
where ${\bf C}$ is complex numbers field.
Since $\cal{H}(\omega)$ is not hermitean, therefore quantum system is
dissipative system. Ginibre ensemble of random matrices
is one of many 
Gaussian Random Matrix ensembles GRME.
The above approach is an example of Random Matrix theory RMT
\cite{Haake 1990,Guhr 1998,Mehta 1990 0}.
The other RMT ensembles are for example
Gaussian orthogonal ensemble GOE, unitary GUE, symplectic GSE,
as well as circular ensembles: orthogonal COE,
unitary CUE, and symplectic CSE.
The distributions of the eigenenergies 
$Z_{1}, ..., Z_{N}$
for $N \times N$ Hamiltonian matrices is given by Jean Ginibre's formula 
\cite{Haake 1990,Guhr 1998,Ginibre 1965,Mehta 1990 1}:
\begin{eqnarray}
& & P(z_{1}, ..., z_{N})=
\label{Ginibre-joint-pdf-eigenvalues} \\
& & =\prod _{j=1}^{N} \frac{1}{\pi \cdot j!} \cdot
\prod _{i<j}^{N} \vert z_{i} - z_{j} \vert^{2} \cdot
\exp (- \sum _{j=1}^{N} \vert z_{j}\vert^{2}),
\nonumber
\end{eqnarray}
where $z_{i}$ are complex-valued sample points
($z_{i} \in {\bf C}$).
For Ginibre ensemble we define complex-valued spacings
$\Delta^{1} Z_{i}$ and second differences $\Delta^{2} Z_{i}$:
\begin{equation}
\Delta^{1} Z_{i}=Z_{i+1}-Z_{i}, i=1, ..., (N-1),
\label{first-diff-def}
\end{equation}
\begin{equation}
\Delta ^{2} Z_{i}=Z_{i+2} - 2Z_{i+1} + Z_{i}, i=1, ..., (N-2).
\label{Ginibre-second-difference-def}
\end{equation}
The $\Delta^{2} Z_{i}$ are extensions of
real-valued second differences
\begin{equation}
\Delta^{2} E_{i}=E_{i+2}-2E_{i+1}+E_{i}, i=1, ..., (N-2),
\label{second-diff-def}
\end{equation}
of adjacent ordered increasingly real-valued energies $E_{i}$
defined for
GOE, GUE, GSE, and Poisson ensemble PE
(where Poisson ensemble is composed of uncorrelated
randomly distributed eigenenergies)
\cite{Duras 1996 PRE,Duras 1996 thesis,Duras 1999 Phys,Duras 1999 Nap,Duras 1996 APPB,Duras 1997 APPB}.

There is an analogy of Coulomb gas of 
unit electric charges pointed out by Eugene Wigner and Freeman Dyson.
A Coulomb gas of $N$ unit charges moving on complex plane (Gauss's plane)
{\bf C} is considered. The vectors of positions
of charges are $z_{i}$ and potential energy of the system is:
\begin{equation}
U(z_{1}, ...,z_{N})=
- \sum_{i<j} \ln \vert z_{i} - z_{j} \vert
+ \frac{1}{2} \sum_{i} \vert z_{i}^{2} \vert. 
\label{Coulomb-potential-energy}
\end{equation}
If gas is in thermodynamical equilibrium at temperature
$T= \frac{1}{2 k_{B}}$ 
($\beta= \frac{1}{k_{B}T}=2$, $k_{B}$ is Boltzmann's constant),
then probability density function of vectors of positions is 
$P(z_{1}, ..., z_{N})$ Eq. (\ref{Ginibre-joint-pdf-eigenvalues}).
Complex eigenenergies $Z_{i}$ of quantum system 
are analogous to vectors of positions of charges of Coulomb gas.
Moreover, complex-valued spacings $\Delta^{1} Z_{i}$
are analogous to vectors of relative positions of electric charges.
Finally, complex-valued
second differences $\Delta^{2} Z_{i}$ 
are analogous to
vectors of relative positions of vectors
of relative positions of electric charges.

The $\Delta ^{2} Z_{i}$ have their real parts
${\rm Re} \Delta ^{2} Z_{i}$,
and imaginary parts
${\rm Im} \Delta ^{2} Z_{i}$, 
as well as radii (moduli)
$\vert \Delta ^{2} Z_{i} \vert$,
and main arguments (angles) ${\rm Arg} \Delta ^{2} Z_{i}$.

\section{Second Difference Distributions}
\label{sec-second-difference-pdf}
We define following random variables
for $N$=3 dimensional Ginibre ensemble:
\begin{equation}
Y_{1}=\Delta ^{2} Z_{1}, 
A_{1}= {\rm Re} Y_{1}, B_{1}= {\rm Im} Y_{1},
\label{Ginibre-Y1A1B1-def}
\end{equation}
\begin{equation}
R_{1} = \vert Y_{1} \vert, \Phi_{1}= {\rm Arg} Y_{1},
\label{Ginibre-polar-second-diff-def}
\end{equation}
and for the generic $N$-dimensional Ginibre ensemble
\cite{Duras 2000 JOptB}:
$W_{1}=\Delta ^{2} Z_{1}$.

Their distributions for 
are given by following formulae 
\cite{Duras 2000 JOptB}:
\begin{eqnarray}
& & f_{Y_{1}}(y_{1})=f_{(A_{1}, B_{1})}(a_{1}, b_{1})=
\label{Ginibre-marginal-pdf-Y1-def} \\
& & =\frac{1}{576 \pi} [ (a_{1}^{2} + b_{1}^{2})^{2} + 24]
\cdot \exp (- \frac{1}{6} (a_{1}^{2}+a_{2}^{2})).
\nonumber
\end{eqnarray}
\begin{equation}
f_{A_{1}}(a_{1})=
\frac{\sqrt{6}}{576 \sqrt{\pi}} (a_{1}^{4}+6a_{1}^{2}+ 51)
\cdot \exp (- \frac{1}{6} a_{1}^{2}),
\label{Ginibre-marginal-pdf-Y1Re-def}
\end{equation}
\begin{equation}
f_{B_{1}}(b_{1})=
\frac{\sqrt{6}}{576 \sqrt{\pi}} (b_{1}^{4}+6b_{1}^{2}+ 51)
\cdot \exp (- \frac{1}{6} b_{1}^{2}),
\label{Ginibre-marginal-pdf-Y1Im-def}
\end{equation}
\begin{eqnarray}
& & f_{R_{1}}(r_{1})=
\label{Ginibre-polar-second-diff-result} \\
& & \Theta(r_{1}) \frac{1}{288}r_{1}(r_{1}^{4}+24) \cdot \exp(- \frac{1}{6} r_{1}^{2}),
\nonumber \\
& & f_{\Phi_{1}}(\phi_{1})= \frac{1}{2 \pi}, \phi_{1} \in [0, 2 \pi].
\nonumber
\end{eqnarray}
\begin{eqnarray}
& & P_{3}(w_{1})=
\label{W1-pdf-I-result} \\
& & = \pi^{-3} \sum_{j_{1}=0}^{N-1} \sum_{j_{2}=0}^{N-1} \sum_{j_{3}=0}^{N-1}
\frac{1}{j_{1}!j_{2}!j_{3}!}I_{j_{1}j_{2}j_{3}}(w_{1}),
\nonumber \\
& & I_{j_{1}j_{2}j_{3}}(w_{1})=
\label{W1-pdf-I-F} \\
& & = 2^{-2j_{2}} 
\frac{\partial^{j_{1}+j_{2}+j_{3}}}
{\partial^{j_{1}} \lambda_{1} \partial^{j_{2}} \lambda_{2}
\partial^{j_{3}} \lambda_{3}}
F(w_{1},\lambda_{1},\lambda_{2},\lambda_{3}) \vert _{\lambda_{i}=0},
\nonumber
\end{eqnarray}
\begin{eqnarray}
& & F(w_{1},\lambda_{1},\lambda_{2},\lambda_{3})=
\label{W1-pdf-I-F-final} \\
& & = A(\lambda_{1},\lambda_{2},\lambda_{3})
\exp[-B(\lambda_{1},\lambda_{2},\lambda_{3}) \vert w_{1} \vert^{2}],
\nonumber
\end{eqnarray}
\begin{eqnarray}
& & A(\lambda_{1},\lambda_{2},\lambda_{3})=
\label{W1-pdf-I-A} \\
& & =\frac{(2\pi)^{2}}
{(\lambda_{1}+\lambda_{2}-\frac{5}{4}) 
\cdot (\lambda_{1}+\lambda_{3}-\frac{5}{4})-(\lambda_{1}-1)^{2}},
\nonumber \\
& & B(\lambda_{1},\lambda_{2},\lambda_{3})=
\label{W1-pdf-I-B} \\
& & =(\lambda_{1}-1) \cdot \frac{2 \lambda_{1}-\lambda_{2}-\lambda_{3}+\frac{1}{2}}
{2 \lambda_{1}+\lambda_{2}+\lambda_{3}-\frac{9}{2}}.
\nonumber
\end{eqnarray}

\section{Conclusions}
\label{sect-conclusions}
We compare second difference distributions for different ensembles 
by defining following dimensionless second differences:
\begin{equation}
C_{\beta} = \frac{\Delta^{2} E_{1}}{<S_{\beta}>},
\label{rescaled-second-diff-GOE-GUE-GSE-PE}
\end{equation}
\begin{equation}
X_{1}=\frac{A_{1}}{<R_{1}>},
\label{Ginibre-X1-def} 
\end{equation}
where $<S_{\beta}>$ are
the mean values of spacings 
for GOE(3) ($\beta=1$),
for GUE(3) ($\beta=2$),
for GSE(3) ($\beta=4$), for PE ($\beta=0$)
\cite{Duras 1996 PRE,Duras 1996 thesis,Duras 1999 Phys,Duras 1999 Nap,Duras 1996 APPB,Duras 1997 APPB},
and $<R_{1}>$ is mean value of radius $R_{1}$ 
for $N$=3 dimensional Ginibre ensemble \cite{Duras 2000 JOptB}.

On the basis of comparison of results for
Gaussian ensembles, Poisson ensemble, and Ginibre ensemble
we formulate homogenization law
\cite{Duras 1996 PRE,Duras 1996 thesis,Duras 1999 Phys,Duras 1999 Nap,Duras 1996 APPB,Duras 1997 APPB,Duras 2000 JOptB}: 
{\em Eigenenergies for Gaussian ensembles, for Poisson ensemble,
and for Ginibre ensemble tend to be homogeneously distributed.}
The second differences' distributions assume global maxima at origin
for above ensembles.
For Coulomb gas 
the vectors of relative positions of vectors
of relative positions of charges statistically vanish.
It can be called stabilisation
of structure of system of electric charges. 

\section{Acknowledgements}
\label{sect-acknowledgements}
It is my pleasure to most deeply thank Professor Jakub Zakrzewski
for formulating the problem.

\end{document}